\documentclass[10pt, conference, compsocconf]{article}

\usepackage[utf8]{inputenc}
\usepackage{empheq}
\usepackage{amsmath,amstext,amsfonts,amssymb}
\usepackage{amsthm}
\usepackage{mathrsfs}
\usepackage{fullpage}
\usepackage{tikz}
\usepackage{tkz-linknodes}
\usepackage{boxedminipage}
\usetikzlibrary{arrows}
\usetikzlibrary{decorations.pathreplacing}
\usetikzlibrary{patterns}
\usepackage{nopageno}
\usepackage{array}
\usepackage{supertabular}
\usepackage{tabularx}

\usepackage{tikz}
\usepackage{tkz-linknodes}
\usepackage{boxedminipage}
\usetikzlibrary{arrows}
\usetikzlibrary{decorations.pathreplacing}
\usetikzlibrary{patterns}
\usetikzlibrary{calc}
\usepackage{scalefnt}
\usepackage{mathrsfs}
\usepackage{subfig}
\usepackage{cases}

\newtheorem{lemma}{\textbf{Lemma}}

\usepackage[textwidth=17mm]{todonotes}
\newboolean{WIP}
\gdef\ifwip{\ifthenelse{\boolean{WIP}}}
\setboolean{WIP}{true}%
\ifwip{
  \newcommand{\indication}[1]{{\it\color{red}#1}\\}
  \newcommand{\Amina}[2][inline]{\todo[color=green!50,#1]{\sf
      \textbf{Notes: } #2}}
}{
  \newcommand{\indication}[1]{}
  \newcommand{\Amina}[2][]{}
} 

\begin{document}
%
\title{Minimizing Energy Consumption of MPI Programs in Realistic Environment}


\author{Amina Guermouche, Nicolas Triquenaux, Beno\^it Pradelle and William Jalby\\\small{Universit\'e de Versailles Saint-Quentin-en-Yvelines}}

\date{}



\maketitle

\begin{abstract}
Dynamic voltage and frequency scaling proves to be an efficient way of
reducing energy consumption of servers. Energy savings are typically
achieved by setting a well-chosen frequency during some program
phases. However, determining suitable program phases and their
associated optimal frequencies is a complex problem. Moreover,
hardware is constrained by non negligible frequency transition
latencies. Thus, various heuristics were proposed to determine and
apply frequencies, but evaluating their efficiency remains an issue.

In this paper, we translate the energy minimization problem into a
mixed integer program that specifically models realistic hardware
limitations. The problem solution then estimates the minimal energy
consumption and the associated frequency schedule. The paper provides
two different formulations and a discussion on the feasibility of each
of them on realistic applications.

\end{abstract}

\section{Introduction}

For a very long time, computing performance was the only metric
considered when launching a program. Scientists and users only cared
about the time it took for a program to finish. Though still often
true, the priority of many hardware architects and system
administrators has shifted to caring more and more about energy
consumption. Solutions reducing the energy enveloppe have been put
forth.

Among the different existing techniques, Dynamic Voltage and Frequency
Scaling (DVFS) proved to be an efficient way to reduce processor
energy consumption. The processor frequency is adapted according to
its workload: When the frequency is lowered without increasing the
execution time, the power consumption and energy are reduced.

With parallel applications in general, and more precisely with MPI
applications, reducing frequency on one processor may have a dramatic
impact on the execution time of the application: Reducing processor
frequency may delay a message sending, and maybe its reception. This
may lead to cascading delays increasing the execution time. To save
energy with respect to application deadline, two main solutions exist:
online tools and offline scheduling. The former try to provide the
frequency schedule during the execution whereas the latter provide it
after an offline study. They both require the application task graph
(either through a previous execution or by focusing on iterative
applications). 

Many online tools~\cite{1559985,Rountree:2009:AMD:1542275.1542340}
identify the \emph{critical path}: the longest path through the graph,
and focus on processors that do not execute these tasks. Typically,
when waiting for a message, the processor frequency is set to the
minimal frequency until the message
arrives~\cite{Rountree:2009:AMD:1542275.1542340}. Although online
tools allow some energy savings, they provide suboptimal
energy saving because of a lack of application knowledge. 

On the other hand, offline scheduling
algorithms~\cite{CPE:CPE2889,5348808} provide the best frequency
execution of each task. 
However, none of the existing algorithms consider most current
multi-core architectures characteristics: (i) cores within the same
processor share the same frequency~\cite{IntelXeonDS} and (ii)
switching frequency requires some time~\cite{MLP13}. 

This paper presents two models based on linear programming which find
the execution frequencies of each task while taking into account the
mutlicore architecture constraints and characteristics
(section~\ref{sec:mip}) previously described. Moreover, we allow the
execution time to be increased if this leads to more energy
savings. The user provides a maximum performance degradation that she
can tolerate. The presented models provide optimal frequency schedule
which minimizes the energy consumption. However, when considering
large applications and large machines, no current solver can provide a
result, even parallel ones. The reason behind this issue is discussed
in section~\ref{sec:mip}.

\section{Context and execution model}\label{sec:context}
We consider MPI applications running on a multi-node platform. The
targeted architectures consider the following characteristics: (i) the
latency of frequency switching is not negligible and (ii) cores within
the same processor share the same frequency. 

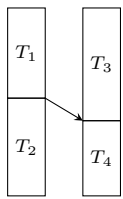
\begin{figure}[h]
\centering
\begin{tikzpicture}
\draw (0,3) rectangle (0.5,1.8) node [midway,font=\scriptsize]{$T_1$};
\draw (0,1.8) rectangle (0.5,0.5) node [midway,font=\scriptsize]{$T_2$};

\draw (1,3) rectangle (1.5,1.5) node [midway,font=\scriptsize]{$T_3$};
\draw (1,1.5) rectangle (1.5,0.5) node [midway,font=\scriptsize]{$T_4$};


\draw [>=stealth,->] (0.5,1.8) -- (1,1.5);
\end{tikzpicture}
\caption{Task graph}\label{fig:taskGraph}
\end{figure}

A process, running on every core, executes a set of tasks. A task,
denoted $T_i$, is defined as the computations between two
communications. The application execution is represented as task graph
where tasks are vertices and edges are messages between the
tasks. Figure~\ref{fig:taskGraph} is an example of the task graph
running on two processes. One process executes tasks $T_1$ and $T_2$
while the other one executes tasks $T_3$ and $T_4$.

\begin{figure}[h]
\centering
\begin{tikzpicture}

\node at (0.75, 3.5) [font=\scriptsize]{$0$};
\node at (2.25, 3.5) [font=\scriptsize]{$1$};
\draw (0,3) rectangle (0.5,1.8) node [midway,font=\scriptsize]{$T_1$};
\draw (0,1.8) rectangle (0.5,0.5) node [midway,font=\scriptsize]{$T_2$};

\draw (1,3) rectangle (1.5,1.5) node [midway,font=\scriptsize]{$T_3$};
\draw (1,1.5) rectangle (1.5,0.5) node [midway,font=\scriptsize]{$T_4$};

\draw (2,3) rectangle (2.5,1.5) node [midway,font=\scriptsize]{$T_5$};
\draw (2,1.5) rectangle (2.5,0.5) node [midway,font=\scriptsize]{$T_6$};


\draw [>=stealth,->] (0.5,1.8) -- (1,1.5);
\draw [>=stealth,->] (2,1.5) -- (1.5,0.5);

\draw [dashed] (-0.5, 2.5) -- (1.5,2.5);
\draw [dashed,red] (-0.75, 2.6) -- (1.5,2.6);
\draw [dashed] (-0.5, 2) -- (1.5,2);
\draw [dashed] (-0.5, 1.2) -- (1.5,1.2);
\node at (-0.25,2.75) [font=\scriptsize]{$f_1$};
\node at (-1.5,2.6)[font=\scriptsize]{request $f_2$};
\node at (-0.25,2.25) [font=\scriptsize]{$f_2$};
\node at (-0.25,1.6) [font=\scriptsize]{$f_3$};
\node at (-0.25,0.85) [font=\scriptsize]{$f_1$};

\draw [dashed] (2, 2.25) -- (3,2.25);
\draw [dashed] (2, 1) -- (3,1);
\node at (2.75,2.625) [font=\scriptsize]{$f_3$};
\node at (2.75,1.625) [font=\scriptsize]{$f_1$};
\node at (2.75,0.75) [font=\scriptsize]{$f_2$};

\end{tikzpicture}
\caption{Frequency switch latency \protect\footnotemark}\label{fig:explanation}
\end{figure}
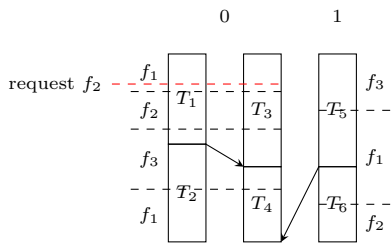

Before going into more details on the execution model, let us provide
an example of the problem we want to solve. Consider the example
provided in Figure~\ref{fig:explanation}. The application is executed
on 3 cores, 2 in the same processor and one in another processor.
Tasks $T_1$, $T_2$, $T_3$ and $T_4$ are executed on processor $0$
while tasks $T_5$ and $T_6$ are executed on processor $1$. In order to
minimize the energy consumption through DVFS, we make the same
assumption as~\cite{5348808}: tasks may have several phases and each
phase can be executed at a specific frequency. Typically on
Figure~\ref{fig:explanation}, task $T_1$ is divided into $3$
phases. The first one is executed at frequency $f_1$, the second one
at frequency $f_2$ and the last one at frequency $f_3$. 

As stressed out before, setting a frequency takes some time. In other
words, when a frequency is requested, it is not set immediately. Thus,
on Figure~\ref{fig:explanation}, when frequency $f_2$ is requested, it
is set some time after. One needs to be careful of such situations
since a frequency may be set after the task which it was requested from
is over.

Moreover, cores within the same processor run at the same
frequency. Hence, on Figure~\ref{fig:explanation}, when $f_1$ is first
set on processor $0$, all the tasks being executed at this time ($T_1$
and $T_3$) are executed at frequency $f_1$. $T_5$ is not affected
since it is on another processor. To provide the best frequency to
execute each task portion, we need to consider all parallel tasks
which are executed at the same time on the processor.

\footnotetext{Note that only the
    latency of the first request is
    represented}

Our model requires the task graph to be provided (through profiling or
a complete execution of the application). Thus, we consider
deterministic applications: for the same parameters and the same input
data, the same task graph is generated. In order to guarantee that
edges are the same over all possible executions, one has to make sure
that the communications between the processes are the same. Non
deterministic communications in MPI are either receptions from an
unknown source (by using \emph{MPI\_Any\_Source} in the reception
call), or non-deterministic completion events (\emph{MPI\_Waitany} for
instance). Any application with such events is considered as
non-deterministic, thus falls out of the scope of the proposed
solution.


\begin{figure}[h]
\centering
\begin{tikzpicture}
\draw (0,3) rectangle (0.5,1.5) node [midway,font=\scriptsize]{$T_1$};
\draw (0,1.5) rectangle (0.5,0.5) node [midway,font=\scriptsize]{$T_2$};

\draw (1,3) rectangle (1.5,2) node [midway,font=\scriptsize]{$T_3$};
\draw (1,1.3) rectangle (1.5,0.5) node [midway,font=\scriptsize]{$T_4$};

\draw (1,2) [dashed] rectangle (1.5,1.3);

\draw [>=stealth,->] (0.5,1.5) -- (1,1.3);
\draw [decoration={brace}, decorate] (1.7,2) [-] -- (1.7,1.3) node [black,midway, right,font=\scriptsize]{$slack$};
\end{tikzpicture}
\caption{Slack time}\label{fig:slack}
\end{figure}
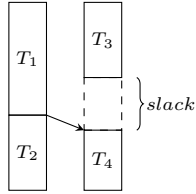

Tasks within a core are totally ordered. If a task $T_i$ ends with a
send event, then the following task $T_j$ starts exactly at the end of
$T_i$. On Figure~\ref{fig:taskGraph}, task $T_2$ starts exactly after
$T_1$ ends. On the other hand, when a task is created by a message
reception ($T_4$ on Figure~\ref{fig:taskGraph}), it cannot start
before all the tasks it depends on finish ($T_1$ and $T_3$) and it has
to wait for the message to be received. If the message arrives after
the end of the task which is supposed to receive it, the time between
the end of the task and the reception is known as \emph{slack} time. On
Figure~\ref{fig:slack}, tasks $T_1$ sends a message to $T_3$ but $T_3$
ends before receiving the messages creating the slack represented by
dotted lines.



A task energy consumption $E_i$ is defined as the product of its
execution time $exec_i$ and its power consumption $P_i$. Since the
application is composed of several tasks, its energy consumption can
be expressed as the sum of the energy consumption of all the
tasks. Thus, the goal translates into providing the set of frequency
to execute each task. 
Hence, one can calculate the application energy consumption as:
\begin{equation}
E=\sum\limits_{i}(E_i) = \sum\limits_{i} (exec_i \times P_i) \label{eq:energy}
\end{equation}

Minimizing the energy consumption of the application is equivalent to
minimizing $E$ in equation~\eqref{eq:energy}. 


For each task $T_i$, both $exec_i$ and $P_i$ depend the frequency of
the different phases of the task. In addition, tasks are not
independent since when executed in parallel on the same processor, the
tasks share the same frequency. Moreover, the overall execution time
of the application depends on all the $exec_i$ and the slack time. To
minimize the energy consumption while still controlling the overall
execution time, we express the problem as a linear program.

\section{Building the linear program}\label{sec:mip}
The following paragraphs describe how the energy minimizatoin problems
translates into a linear programming. We first describe the precedence
constraints between the tasks, then we describe two formulations which
consider the architecture constraints. Finally, we discuss the
feasibility of the described solutions.

\subsection{Precedence constraints}

Let $T_i$ be a task defined by its start time $bT_i$ and its end time
$eT_i$. The beginning of tasks is bounded by the precedence relation
between them. As already stressed out, a task cannot start before its
direct predecessors complete their execution. As explained in
section~\ref{sec:context}, if $T_i$ sends a message, its child task
$T_j$ starts exactly when $T_i$ ends since the end of the
communication means the beginning of the next task. 
This translates to:
\begin{equation*}
bT_j = eT_i
\end{equation*}

\begin{table}
\begin{center}
\begin{tabular}{|l|l|}
\hline
$bT_i$& Beginning of a task $T_i$\\
$eT_i$& End of a task $T_i$\\
$bTs_i$& Beginning of a slack task $Ts_i$\\
$eTs_i$& End of a slack task $Ts_i$\\
$exec_i^f$& The execution time of a task $T_i$ if executed completely at frequency $f$\\
$tT_i^f$& The time during which the task $T_i$ is executed at frequency $f$\\
$\delta_i^f$& The fraction of time a task $T_i$ spends at frequency $f$\\
$M_j^i$& Message transmission time from task $T_j$ to task $T_i$\\
\hline
\end{tabular}\caption{Task variables}
\end{center}
\end{table}

On the other hand, when $T_i$ ends with a message reception from
$T_k$, one has to make sure that its successor task $T_j$ starts after
both tasks end. Moreover, as pointed out in section~\ref{sec:context},
when a task receives a message, some slack may be introduced before
the reception. Slack is handled the same way tasks are: it has a
start and an end time and it can be executed at different frequencies
depending on the tasks on the other cores. On Figure~\ref{fig:slack},
the slack after $T_3$ may be executed at different frequencies whether
it is executed in parallel with $T_1$ or $T_2$.

To ease the presentation, we assume that each task $T_i$ receiving a
message (from a task $T_k$) is followed by a slack task, denoted
$Ts_i$. The beginning of $Ts_i$, denoted $bTs_i$ is exactly equal to
the end of $T_i$,
\begin{equation}
bTs_i = eT_i\label{eq:slack_start}
\end{equation}
whereas its end time, denoted $eTs_i$, is at least equal to the
arrival time of the message from $T_k$. Let $M_k^i$ denote the
transmission time from $T_k$ to $T_i$. Thus:
\begin{equation}
eTs_i \geq eT_k + M_k^i\label{eq:slack_message}
\end{equation}

Note that a task may receive messages from different processes (after
a collective communication for example) and
equation~\ref{eq:slack_message} has to be valid for all of them.

Finally, since $T_j$, the successor task of $T_i$ has to start after
$T_i$ and $T_k$ finish, one just needs to make sure that:
\begin{equation*}
bT_j =eTs_i
\end{equation*}


In order to compute the end time of a task $T_i$ ($eT_i$), one has to
evaluate the execution time of $T_i$. As explained above, a task may
be executed at different frequencies. Let $exec_i^f$ be the execution
time of $T_i$ if executed completely at frequency $f$. Every frequency
can be used to run a fraction $\delta_i^f$ of the total execution of
the task. Let $tT_i^f$ be the fraction of time $T_i$ spends at
frequency $f$. It can be expressed as: $tT_i^f = \delta_i^f \times
exec_i^f$. Thus, the end time of a task is:

\begin{equation*}
eT_i = bT_i + \sum\limits_{f} tT_i^f
\end{equation*}

Note that one has to make sure that a task is completely executed:
\begin{equation}
\sum\limits_{f} \delta_i^f = 1
\label{eq:task_completed}
\end{equation}

Finally, since the power consumption depends on the frequency, let
$P_i^f$ be the power consumption of the task $T_i$ when executed at
frequency $f$. Using this formulation, the objective function of the
linear program becomes:

\begin{equation}
min (\sum\limits_{T_i} (\sum\limits_f (tT_i^f \times P_i^f)))\label{eq:objective}
\end{equation}



One can just use $tT_i^f$ in the objective function as it is expressed
in equation~\eqref{eq:objective}, and the solver would provide the
values of $tT_i^f$ of all tasks at all frequencies. This solution was
presented in~\cite{5348808}. The provided solution can be used on
different architectures than the ones we target in this work. As a
matter of fact, nothing constrains parallel tasks on one processor to
run at the same frequency, and the threshold of switching frequency is
not considered either. Moreover, no constraint on the execution time
is expressed. The following paragraphs first describe how the
performance is handled then they introduce additional constraints the
handle the architecture constraints and execution time.

\subsection{Execution time constraints}
The performance of an application is a major concern; whether the
energy consumption is considered or not. In this paragraph we provide
constraints which consider the execution time of the application.
In MPI, all programs end with \emph{MPI\_Finalize} which is similar to
a global barrier. Let $last\_task^i$ be the last task on core $i$ (the
\emph{MPI\_Finalize} task). Since the application ends with a global
communication, every task $last\_task^i$ is followed by a slack task
$last\_slack\_task^i$. The difference between the global communication
slack and the other slack tasks lies in the end time: the end time of
all slack tasks of a global communication is the same (all processes
leave the barrier at the same time). Thus, for every couple of cores
$(i, j)$:
\begin{equation}
elast\_slack\_task^i = elast\_slack\_task^j
\end{equation}

Let $total\_Time$ be the application execution time: It is equal to
the end time of the last slack task.
\begin{equation}
total\_Time = elast\_slack\_task^i
\end{equation}

However, in some cases, increasing the execution time of an
application could benefit to energy consumption. In order to allow
this performance loss to a specified extent, the user limits the
degradation to a factor $x$ of the maximal performance. 
Let $exec\_Time$ be the execution time when all tasks run at the
maximal frequency, and $x$ the maximum performance loss percentage
allowed by the user. The following constraint allows performance loss
with respect to $x$:

\begin{equation*}
total\_Time \leq exec\_Time + \frac{exec\_Time \times x}{100}
\end{equation*}

The next sections describe two different formulations. In the first
formulation, the solver is provided with all possible task
configurations and chooses the one minimizing energy
consumption. 
In the second formulation, the solver provides the exact time of every
frequency switch on each processor.

\subsection{Architecture constraints: the workload approach}\label{ssec:workloads}

In order to provide the optimal frequency schedule, the linear program
is provided with all possible task configurations, \emph{i.e.}, all
possible of parallel tasks, known as workloads. Then the solver
provides the execution frequency of each workload.

\subsubsection{Shared frequency constraint}
\begin{figure}[h]
\centering
\subfloat[f\_max\label{fig:workload_f_max}]{
\begin{tikzpicture}





\draw (0,3) rectangle (0.5,1.5) node [midway,font=\scriptsize]{$T_1$};
\draw (0,1.5) [dashed] rectangle (0.5,1) node [midway,font=\scriptsize]{$Ts_1$};
\draw (0,1) rectangle (0.5,0) node [midway,font=\scriptsize]{$T_2$};

\draw (1,3) rectangle (1.5,2) node [midway,font=\scriptsize]{$T_3$};
\draw (1,2) [dashed] rectangle (1.5,1) node [midway,font=\scriptsize]{$Ts_3$};
\draw (1,1) rectangle (1.5,0) node [midway,font=\scriptsize]{$T_4$};

\draw [>=stealth,->] (-0.5,3) -- (0,1);
\draw [>=stealth,->] (2,3) -- (1.5,1);

\draw [red,dashed] (0,2) -- (2,2);
\draw [red,dashed] (0,1.5) -- (2,1.5);
\draw [red,dashed] (0,0) -- (2,0);
\draw [red,dashed] (0,1) -- (2,1);



\end{tikzpicture}
}
\hspace{2cm}
\subfloat[f\_min\label{fig:workload_f_min}]{
\begin{tikzpicture}
\draw (0,3) rectangle (0.5,1.25) node [midway,font=\scriptsize]{$T_1$};
\draw (0,1.25) [dashed] rectangle (0.5,1) node [midway,font=\scriptsize]{$Ts_1$};
\draw (0,1) rectangle (0.5,0) node [midway,font=\scriptsize]{$T_2$};

\draw (1,3) rectangle (1.5,1) node [midway,font=\scriptsize]{$T_3$};
\draw (1,1) rectangle (1.5,0) node [midway,font=\scriptsize]{$T_4$};

\draw [>=stealth,->] (-0.5,3) -- (0,1);
\draw [>=stealth,->] (2,3) -- (1.5,1);

\draw [red,dashed] (-0.5,1.25) -- (2,1.25);
\draw [red,dashed] (-0.5,0) -- (2,0);
\draw [red,dashed] (-0.5,1) -- (2,1);





\end{tikzpicture}
}
\caption{Workloads}\label{fig:workload}
\end{figure}
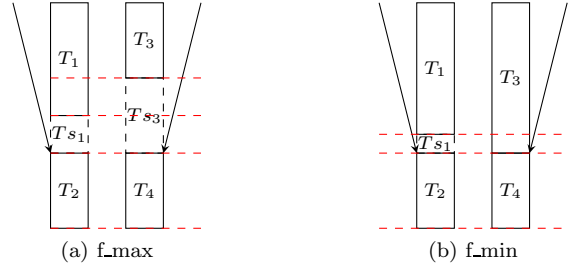

We need to express that tasks executed at the same time on the same
processor run at the same frequency. Hence, we first need to identify
tasks executed in parallel on the same processor. 
Depending on the frequency being used, the set of parallel tasks may
change. Figure~\ref{fig:workload} is an example of two different
executions running at the maximal and minimal frequency. Only
processes that belong to the same processor are represented. In
Figure~\ref{fig:workload_f_max}, when the processor runs at $f\_max$,
the set of couple of tasks which are parallel is: $\{(T_1,T_3), (T_1,
Ts_3), (Ts_1, Ts_3), (T_2, T_4)\}$ (represented by red dotted
lines). When the frequency is set to $f\_min$
(Figure~\ref{fig:workload_f_min}), the slack after $T_3$ is completely
covered and the set of parallel tasks becomes: $\{(T_1, T_3), (Ts_1,
T_3), (T_2, T_4)\}$.

In order to provide all possible configurations, we define the
processor workloads. A workload, denoted $W_i$ is tuple of potentially
parallel tasks. In Figure~\ref{fig:workload}, $W_1 = (T_1,T_3)$, $W_2
= (Ts_1, T_3)$, $W_3 = (T_1, Ts_3)$ represent a subset of the possible
workloads. Note that there are no workloads with the same set of
tasks. In other words, once a task in a workload is over, a new
workload begins. On the other hand, a task can belong to several
workloads (like $T_1$ in Figure~\ref{fig:workload_f_max}).


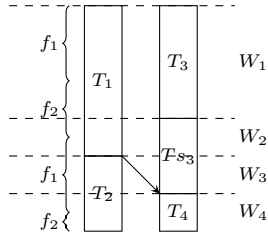
\begin{figure}
\centering
\begin{tikzpicture}
\draw (0,3) rectangle (0.5,1) node [midway,font=\scriptsize]{$T_1$};
\draw (0,1) rectangle (0.5,0) node [midway,font=\scriptsize]{$T_2$};
\draw (1,3) rectangle (1.5,1.5) node [midway,font=\scriptsize]{$T_3$};
\draw (1,1.5) rectangle (1.5,0.5) node [midway,font=\scriptsize]{$Ts_3$};
\draw (1,0.5) rectangle (1.5,0) node [midway,font=\scriptsize]{$T_4$};
\draw [>=stealth,->] (0.5,1) -- (1,0.5);
\draw [decoration={brace,mirror}, decorate] (-0.2,3) [-] -- (-0.2,2) node [black,midway, left,font=\scriptsize]{$f_1$};
\draw [decoration={brace,mirror}, decorate] (-0.2,2) [-] -- (-0.2,1.25) node [black,midway, left,font=\scriptsize]{$f_2$};
\draw [decoration={brace,mirror}, decorate] (-0.2,1.25) [-] -- (-0.2,0.25) node [black,midway, left,font=\scriptsize]{$f_1$};
\draw [decoration={brace,mirror}, decorate] (-0.2,0.25) [-] -- (-0.2,0) node [black,midway, left,font=\scriptsize]{$f_2$};
\draw [-,dashed](-1,3) -- (2,3); 
\node at (2.25,2.25)[font=\scriptsize]{$W_1$};
\draw [-,dashed](-1,1.5) -- (2,1.5); 
\node at (2.25,1.25)[font=\scriptsize]{$W_2$};
\draw [-,dashed](-1,1) -- (2,1); 
\node at (2.25,0.75)[font=\scriptsize]{$W_3$};
\draw [-,dashed](-1,0.5) -- (2,0.5);
\node at (2.25,0.25) [font=\scriptsize]{$W_4$};
\end{tikzpicture}
\caption{Workloads and tasks execution}\label{fig:workload_diff_f}
\end{figure}

\begin{table}
\begin{center}
\begin{tabular}{|l|l|}
\hline
$bW_i$ & Beginning of a workload $W_i$\\
$eW_i$ & End of a workload $W_i$\\
$tW_i^f$ & The time a workload $W_i$ is executed at frequency $f$\\
$dW_i$ & The duration of a workload\\
$\overline{tW_i^f}$ & A binary variable used to say if a workload is executed at a frequency $f$ or not\\
\hline
\end{tabular}\caption{Workload formulation variables}
\end{center}
\end{table}

Recall that our goal in to calculate the fraction of time a tasks
should spend at each frequency ($tT_i^f$) in order to minimize the
energy consumption of the application according to the objective
function~\eqref{eq:objective}. Since tasks may be executed at several
frequencies, so does a workload
. In Figure~\ref{fig:workload_diff_f}, the workload $W_1 = (T_1, T_3)$
is executed at frequency $f_1$ then at frequency $f_2$. Thus, since
$T_1$ belongs to both $W_1 = (T_1, T_3)$ and $W_2= (T_1, Ts_3)$, the
execution time of $T_1$ at frequency $f_1$ ($tT_1^{f_1}$) can be
calculated by using the fraction of time $W_1$ and $W_2$ spend at
frequency $f_1$. In other words, the execution time of a task can be
calculated according to the execution time of the workloads it belongs
to. 
Let $tW_i^f$ be the fraction of time the workload $W_i$ spends at
frequency $f$. Thus:
\begin{equation}
tT_i^f = \sum\limits_{W_j, T_i \in W_j} tW_j^f
\label{eq:tTtW}
\end{equation}
Using the execution
time of a workload at a specific frequency ($tW_i^f$), one can
calculate the duration of a workload, $dW_i$ as:

\begin{center}
\begin{equation*}
dW_i = \sum\limits_{f} tW_i^f \label{const:dw}
\end{equation*}
\end{center}
\subsubsection{Handling frequency switch delay}
Recall that one of the problems when considering DVFS is the time
required to actually set a new frequency. Thus, before setting a
frequency, one has to make sure that duration of the workload is long
enough to tolerate the frequency change since changing frequency takes
some time. In other words, if the frequency $f$ is set in a $W_i$,
$tW_i^f$ is larger than a user-defined threshold, denoted $Th$.
\begin{center}
\begin{equation}
\forall W_i , \forall f: tW_i^f \geq Th \times \overline{tW_i^f} \label{eq:tWifTh}
\end{equation}
\end{center}
$\overline{tW_i^f}$ is a binary variable used to guarantee that
definition~\eqref{eq:tWifTh} remains true when $tW_i^f=0$.
\begin{equation}
\overline{tW_i^f} = 
\begin{cases}
0 & tW_i^f  = 0 \\
1 & otherwise
\end{cases}
\label{eq:twifB}
\end{equation}
The expression of definition~\eqref{eq:twifB} as a mixed binary
programming formulation is expressed in the appendix.

\subsubsection{Valid workload filtering}
The linear program is provided with all possible workloads, then it
provides the different $tW_j^f$ for each workload. However, all
workloads cannot be present in one execution. In
Figure~\ref{fig:workload}, $W_1 = (T_1, Ts_3)$ and $W_2 = (Ts_1, T_3)$
are both possible workloads, but they cannot be in the same execution,
because if $W_1$ is being executed, it means that $T_3$ is over (since
$Ts_3$ is after $T_3$) thus $W_2$ cannot appear later since $Ts_1$ and
$T_3$ are never parallel. Thus, in order to prevent $W_1$ and $W_2$
from both existing in one execution, we just need to check whether the
tasks of the workload can be parallel or not. Two tasks are not
parallel if one ends before the beginning of the second. Since we
consider workloads, we focus only on the beginning and end time of the
workload itself. Let $bW_i$ and $eW_i$ be the start time and the end
time of the workload $W_j = (T_1, \ldots, T_i, \ldots, T_n)$. They are
such that:
\begin{eqnarray}
bW_j & >= & bT_i \label{eq:workload_b}\\
eW_j & <= & eT_i\label{eq:workload_e}
\end{eqnarray}
Note that although the beginning and the end of the workload are not
exactly defined, this definition makes sure that the beginning or the
end of a task start a new workload. Moreover, the complete execution
of a task are guaranteed thanks to equations~\eqref{eq:task_completed}
and~\eqref{eq:tTtW}.

\begin{figure}[h]
\centering
\begin{tikzpicture}
\draw (0,3) rectangle (0.5,1) node [midway,font=\scriptsize]{$T_1$};
\draw (0,1) rectangle (0.5,0.5) node [midway,font=\scriptsize]{$Ts_1$};
\draw (0,0.5) rectangle (0.5,0) node [midway,font=\scriptsize]{$T_2$};
\draw (1,3) rectangle (1.5,1.5) node [midway,font=\scriptsize]{$T_3$};
\draw (1,1.5) rectangle (1.5,0.5) node [midway,font=\scriptsize]{$Ts_3$};

\draw [dashed](0.5,1) -- (2.5,1) node [right,font=\scriptsize]{$bTs_1$ \textcolor{red}{$bW_2 \geq bTs_1$} and \textcolor{red}{$bW_2 \geq bT_3$}. Thus the workload must at least start here};
\draw [dashed](0.5,0.5) -- (2.5,0.5) node [right,font=\scriptsize]{$eTs_1$};
\draw [dashed](1.5,3) -- (2.5,3) node [right,font=\scriptsize]{$bT_3$};
\draw [dashed](1.5,1.5) -- (2.5,1.5) node [right,font=\scriptsize]{$eT_3$ \textcolor{red}{$eW_2 \leq eTs_1$} and \textcolor{red}{$eW_2 \leq eT_3$}. Thus the workload must at most end here};

\end{tikzpicture}
\caption{Negative workload duration for impossible workloads}\label{fig:workload_duration}
\end{figure}
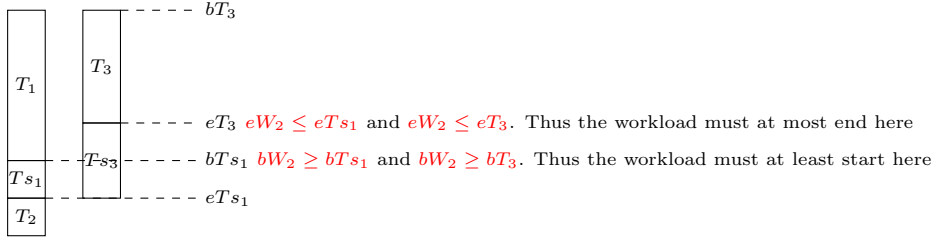

Figure~\ref{fig:workload_duration} is an example of a workload that
cannot exist. Let us assume the execution represented in
Figure~\ref{fig:workload_duration}, and let us focus on the workload
$W_1=(T_1,Ts_3)$. Let us also assume that with other frequencies, a
possible workload is $W_2=(T_3,Ts_1)$. As explained above, $W_1$ and
$W_2$ cannot both exist in the same execution because of precedence
constraints. It is obvious from the example that $T_3$ and $Ts_1$ are
not parallel, let us see how it translates to workloads. Since $W_2$
has to start after both $T_3$ and $Ts_1$ begins, then it starts after
$Ts_1$ (since $bTs_1 \geq bT_3$
Figure~\ref{fig:workload_duration}). The same way it ends before
$eT_3$. But since $eT_3 \leq bTs_1$ (as shown in
Figure~\ref{fig:workload_duration}) then the duration of $W_2$ should
be negative which is not possible.

Thus, we identify workloads which cannot be in the execution as
workloads which end before they begin. The duration of a workload is
such that:
\begin{equation}
dW_i = 
\begin{cases}
0 & eW_i < bW_i \\
eW_i - bW_i & otherwise
\end{cases}
\label{eq:dWi}
\end{equation}

In the appendix (section~\ref{sec:appendix}), we proove that if two
workloads cannot be in the same execution (because of the precedence
constraints), then the duration of at least one of them is $0$
(paragraph~\ref{ssec:workload_proof}).




\subsubsection{Discussion}
The appendix (section~\ref{sec:appendix}) provides a detailed
formulation of the energy minimization problem using workloads. The
formulation shows the use of two binary variables: one to express the
threshold constraint and one to calculate the duration of the
workload. With these two variables, the formulation is not linear
anymore, which requires more time to solve (especially when the number
of workloads is important).

Moreover, we tried providing all possible workloads of one of the NAS
parallel benchmarks on class C on 16 processes (IS.C.16) on a machine
equiped with 16 GB of memory. The application task graph is composed
of 630 tasks. The generated data (\emph{i.e.} the number of workloads)
could not fit in the memory of the machine. Thus, even with no binary
variables, providing all possible workloads is not possible when
considering real applications.

In the following section, we provide another formulation which
requires only the task graph.

\subsection{Architecture constraints: the frequency switch approach}\label{ssec:fswitch}
As explained earlier, our goal is to minimize the energy consumption
of a parallel application using DVFS. In order to do so, we express
the problem as a linear program. We consider that the program is
represented as a task graph and each task can have several phases. The
difficulty of the formulation is to provide, for each task, the
frequency of each of its phases ($tT_i^f$) since one has to make sure
that parallel tasks must run at the same frequency. In this section,
we provide another formulation which considers the time to set a new
frequency on the whole processor instead of considering tasks
independently and then force parallel tasks to run at the same
frequency.


\subsubsection{Frequency switch overhead}
Let $c_{jp}^f$ be the time the frequency $f$ is set on the processor
$p$, $j$ being the sequence number of the frequency switching.
Figure~\ref{fig:cipf} represents the execution of four tasks on two
cores of the same processor $p$. In the example, we assume that there
are only $3$ possible frequencies. The different $c_{jp}^f$ are
numbered such that the minimum frequency $f_1$ corresponds to the
switching time $c_{1p}^{f_1}, c_{4p}^{f_1}, \ldots$, the frequency
$f_2$ corresponds to the frequency changes $c_{2p}^{f_2},
c_{5p}^{f_2}, \ldots$ and so on. A frequency $f_1$ is applied during a
time which can be calculated as
$c_{\{i+1\}p}^{f_2}-c_{ip}^{f_1}$. This can be translated to:
\begin{equation*}
c_{\{i+1\}p}^{f_2} \geq c_{ip}^{f_1}
\end{equation*}
\begin{table}[h]
\centering
\begin{tabular}{|l|l|}
\hline
$c_{ip}^f$ & Time of the $i^{th}$ frequency switch on processor $p$. The frequency $f$ is the one set\\
$d_{ij}^f$ & The amount of time a frequency $f$ is set for the task $i$ for the frequency switch $j$\\
\hline
\end{tabular}\caption{Frequency switch formulation variables}
\end{table}

Note that some frequencies may not
be set if the duration is zero. In figure~\ref{fig:cipf}, frequency
$f_3$ is not set since $c_{31}^{f_3} = c_{41}^{f_1}$.

\subsubsection{Handling frequency switch delay}
As explained earlier, changing frequency takes some time. Thus, for a
change to be applied, its duration has to be longer than the
user-defined threshold $Th$. Let $\zeta_{ip}^f$ be a binary
variable, 
such that:
\begin{equation}
\zeta_{ip}^f = 
\begin{cases}
0 &  c_{\{i+1\}p}^{f'} - c_{ip}^f = 0 \\
1 & otherwise
\end{cases}
\label{eq:threshold_bool}
\end{equation}
The threshold condition can be expressed as:
\begin{equation*}
 c_{\{i+1\}p}^{f'} - c_{ip}^f \geq Th \times \zeta_{ip}^f \label{const:tWifTh}
\end{equation*}
We detail how equation~\eqref{eq:threshold_bool} is translated into
mixed binary programming constraints in the appendix.

\subsubsection{Shared frequency constraints}
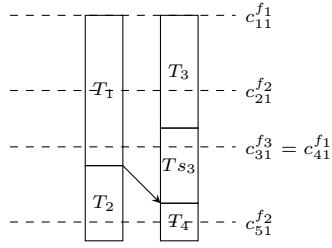
\begin{figure}
\centering
\begin{tikzpicture}
\draw (0,3) rectangle (0.5,1) node [midway,font=\scriptsize]{$T_1$};
\draw (0,1) rectangle (0.5,0) node [midway,font=\scriptsize]{$T_2$};
\draw (1,3) rectangle (1.5,1.5) node [midway,font=\scriptsize]{$T_3$};
\draw (1,1.5) rectangle (1.5,0.5) node [midway,font=\scriptsize]{$Ts_3$};
\draw (1,0.5) rectangle (1.5,0) node [midway,font=\scriptsize]{$T_4$};
\draw [>=stealth,->] (0.5,1) -- (1,0.5);
\draw [-,dashed](-1,3) -- (2,3) node [right,font=\scriptsize]{$c_{11}^{f_1}$};
\draw [-,dashed](-1,2) -- (2,2) node [right,font=\scriptsize]{$c_{21}^{f_2}$};
\draw [-,dashed](-1,1.25) -- (2,1.25) node [right,font=\scriptsize]{$c_{31}^{f_3}=c_{41}^{f_1}$};
\draw [-,dashed](-1,0.25) -- (2,0.25) node [right,font=\scriptsize]{$c_{51}^{f_2}$};
\end{tikzpicture}
\caption{Frequency switches example}\label{fig:cipf}
\end{figure}

Once the threshold condition is satisfied, one can calculate the time
a task spends at each frequency, \emph{i.e} $tT_i^f$, according to
$c_{jp}^f$. On Figure~\ref{fig:cipf}, initially, tasks $T_1$ and $T_3$
run in parallel at frequency $f_1$. The time $T_3$ spends at frequency
$f_1$ is $c_{21}^{f_2} - c_{11}^{f_1}$ whereas $T_1$ is executed twice
at $f_1$. It spends $(c_{21}^{f_2} - c_{11}^{f_1}) + (eT_1 -
c_{41}^{f_1})$ at frequency $f_1$. Let $d_{ij}^f$ be the time the task
$T_i$ spends at frequency $f$ after the frequency switch $j$. Back to
Figure~\ref{fig:cipf}, $d_{11}^{f_1}=c_{21}^{f_2} - c_{11}^{f_1}$ and
$d_{14}^{f_1}= eT_1 - c_{41}^{f_1}$. $tT_1^{f_1}$ becomes
$tT_1^{f_1}=d_{11}^{f_1} + d_{14}^{f_1}$.

The above translates to:
\begin{equation*}
tT_i^f = \sum\limits_{j} d_{ji}^f \\
\end{equation*}
Note that a task is not impacted by a frequency change if it ends
before the change or begins after the next change. In other words,
$d_{ij}^{f_1} = 0$ if $eT_i \leq c_{jp}^{f_1}$ or $bT_i \geq
c_{\{j+1\}p}^{f_2}$. Otherwise, $d_{ij}^{f_1}$ can be calculated as
$min(eT_i, c_{\{j+1\}p}^{f_2}) - max(bT_i, c_{jp}^{f_1})$.
\begin{equation}
d_{ji}^f =
\begin{cases}
0 & eT_i \leq c_{jp}^f \hspace{0.1cm} or \hspace{0.1cm} bT_i \geq c_{\{i+1\}p}^{f'}\\
min (eT_i, c_{\{j+1\}p}^{f'}) - max (bT_i, c_{jp}^f) & otherwise
\end{cases}
\label{eq:dij}
\end{equation}
%
%


\subsection{Discussion}
The appendix (section~\ref{sec:appendix}) provides the complete
formulation of the problem using the frequency switch time
variables. In addition to the binary variable used to satisfy the
frequency switch overhead, for each task and for each frequency
switch, five additionnal binary variables are used. Thus, for $n$
tasks and $m$ frequency switch considered, $5\times n\times m$ binary
variables are required. Mixed integer programming is
NP-hard~\cite{Garey:1990:CIG:574848}, thus, with such a number of
binary variables, no solution can be provided.

When comparing the workload approach and the frequency switch
approach, one can notice that the former needs less binary variables
and should be able to provide results. However, because all possible
workloads have to be provided to the solver, it is as complex because
of the memory required. Thus, if a very large memory is available,
then the workload solution is the one to be used. And if new faster
binary resolution techniques are provided, then the frequency switch
solution should be used.

Several heuristics can be assumed in order to reduce the time to solve
the problem. First, one can consider iterative applications, and solve
the problems for only one iteration then apply it the remaining
ones. However, this solution strongly depends on the number of tasks
per iterations. We tried this solution on some kernels (NAS Parallel
Benchmarks~\cite{NAS}) and the solver could not provide any result
after several hours.

The most promising heuristic is to consider the tasks at the processor
level instead of the core level. Thus, the only architecture
constraint which needs to be considered is the frequency overhead
one. This study is part of our current work and will be discussed in
further studies.

\section{Related Work}\label{sec:related}
DVFS scheduling has been widely used to improve processor energy
consumption during application execution. We focus on studies assuming
a set of dependent tasks represented as a direct acyclic graph (DAG).

A lot of studies tackle task mapping problem while minimizing energy
consumption either with respect to task deadlines~\cite{1012617} or by
trying to minimize the deadline as
well~\cite{DBLP:journals/jips/LeeZ09}. When considering an already
mapped task graph, studies provide the execution speed of each task
depending on the frequency model:
continuous~\cite{DBLP:conf/hipc/BaskiyarP06} or
discrete~\cite{4100345}. Some studies also provide a set of
frequencies to execute a task~\cite{Gruian01lenes:task} (executing a
task at multiple frequencies is known as
VDD-Hopping). In~\cite{CPE:CPE2889}, the authors present a complexity
study of the energy minimization problem depending on the frequency
model (continuous frequencies, discrete frequencies with and without
VDD-Hopping). Finally studies
like~\cite{Mochocki:2002:RVV:774572.774679} and~\cite{1388389}
consider frequency transition overhead. Although these studies should
provide an optimal frequency schedule, they do not consider the
constraints of most current architectures and more specifically the
shared frequency among all cores of the same processor.


When considering linear programming formulation to minimize
application energy consumption, many formulations have been proposed
in the past. When considering single processor, \cite{708188} provides
an integer linear programming formulation with negligible frequency
switching overhead. The same problem but considering frequency
transition overhead was addressed in~\cite{1522778}. The author also
provide a linear-time heuristic algorithm which provides near-optimal
solution.

The work presented in~\cite{5348808} is the closest to the work
presented in this paper. In~\cite{5348808}, the authors present a
linear programming formulation of the minimization energy problem
where tasks can be executed at several frequencies. Both slack energy
and processor energy consumption are considered in the minimization
and a loose deadline is considered. In a similar way, \cite{5562883}
provides a scheduling algorithm and an integer linear programming
formulation of the energy minimization problem on heterogeneous
systems with a fixed deadline. The formulation is very close to the
one described in~\cite{5348808}, but the authors also considered
communication energy consumption. However, they do not consider slack
time and its power consumption when solving the
problem. In~\cite{Liu2014101} the authors use an integer linear
programming formulation of the problem where only task with slack time
are slowed down, whereas other tasks are run at maximal frequency. The
program is used to compute the best frequency execution of a task.

Although previous studies provide different solutions and formulations
for DVFS scheduling, few of them consider current architecture
constraints. While some previous studies consider frequency transition
overhead~\cite{1522778,DBLP:journals/jcse/Kim10}, none of them
consider the fact that cores within the same processor run at the same
frequency. This paper describes a mixed linear programming formulation
that guarantees that parallel tasks on the same processor run at the
same frequency. Moreover, it shows that it is possible to relax the
deadline if it leads to energy saving.

\section{Conclusion}
The goal of this paper was to provide a study on how energy
minimization problem of a parallel execution of an MPI-like program
can be addressed and formulated when considering most current
architecture constraints.  In order to do so, we used linear
programming formulation. Two different formulations were
described. Their goal is to minimize the energy consumption with
respect to a user-defined deadline by providing the optimal frequency
schedule. Both solutions use a number of binary variables which is
proportional to the number of tasks. Used as they are, these
formulations should provide an optimal solution but are costly in
terms of memory and resolution time, despite the use of fast parallel
solvers like gurobi~\cite{gurobi}.

We are currently working on introducing heuristics to relax the
architecture constraints by building tasks on the processor level
instead of the core level. Using such heuristics seems to drastically
reduce the time needed to solve the problem.

\section{Appendix}\label{sec:appendix}
This appendix summarizes the set of constraints of both formulations
described in paragraphs~\ref{ssec:workloads} and
\ref{ssec:fswitch}. 
We start by describing how each non linear constraint which appears in
sections~\ref{ssec:workloads} and~\ref{ssec:fswitch} is expressed. For
a more complete description and explanation, the reader can refer
to~\cite{Bisschop99aimms}.


\subsection{Expressing non linear constraints}
Section~\ref{sec:mip} presents different non continuous variables
(definitions~\ref{eq:twifB}, \eqref{eq:dWi} and \eqref{eq:threshold_bool},
\eqref{eq:dij}). In this section, we briefly explain how this kind of
expressions translates to inequalities using binary variables.

\begin{enumerate}
\item If-then statement with 0-1 variables: Expressing conditions like:
\begin{equation*}
\overline{x} = 
\begin{cases}
0 & x  = 0 \\
1 & otherwise
\end{cases}
\end{equation*}
(for instance, definition~\ref{eq:twifB}) requires the use of a large
constant $M$ such that:
\begin{eqnarray}
x & \leq & M \times \overline{x} \label{eq:bool1}\\
x & \geq & \overline{x} \times \epsilon\label{eq:bool2}
\end{eqnarray}
Thus, when $x = 0$, \eqref{eq:bool2} forces
$\overline{x}$ to be equal to $0$ and when $x \neq 0$,
\eqref{eq:bool1} is used to set the value of
$\overline{x}$ to $1$. 

Note that, equation~\eqref{eq:tWifTh}, which guarantees that $tW_i^f
\geq Th \times \overline{tW_i^f}$ makes \eqref{eq:bool2}
useless (since $Th > \epsilon$). Thus, \eqref{eq:bool2} is
never used in the set of constraints.

\item If-then statement with real variables: Expressing formulas like:
\begin{equation*}
z = 
\begin{cases}
0 & y < x \\
y - x & otherwise
\end{cases}
\end{equation*}
(definition~\eqref{eq:dWi} for instance) is similar to the previous
formulation in the sens that it requires the use of a big constant
$M$. A binary variable $bin$ is used such that when $y - x \leq 0$,
$bin = 0$.
\begin{eqnarray}
y - x & \leq & M\times bin \label{eq:bool3}\\
x - y & \leq & M \times (1-bin)\label{eq:bool4}
\end{eqnarray}

Thus, when $y \leq x$, \eqref{eq:bool3} is always valid
regardless the value of $bin$. Hence, \eqref{eq:bool4} forces
$bin$ to be equal to $0$. Similarly, when $y \geq x$,
equation~\eqref{eq:bool3} forces $bin$ to $1$.

Once $bin$ is defined, $z$ can be expressed as:
\begin{eqnarray}
y - x & \leq & z \leq M\times bin \label{eq:noncontinu1}\\
y - x + z & \leq & 2 \times (y - x) + M \times (1 - bin)\label{eq:noncontinu2}
\end{eqnarray}
Thus, when $y \leq x$, $bin = 0$ (from \eqref{eq:bool3}) and
\eqref{eq:noncontinu1} forces $z$ to be $0$ (since all
variable are positive) and \eqref{eq:noncontinu2} is always
valid. Similarly, when $y \geq x$, $bin = 1$ (from
\eqref{eq:bool4}) and \eqref{eq:noncontinu1}
and \eqref{eq:noncontinu2} become:
\begin{eqnarray*}
y - x & \leq & z \leq M \\
z & \leq & y - x
\end{eqnarray*}
Thus $y - x \leq z \leq y - x$ which makes $z = y - x$.

\item Maximums: Maximums can be expressed by reformulating the definition as:
\begin{equation*}
z  =  max(x, y)
   = x +
\begin{cases}
0 & x \geq y\\
y - x & otherwise
\end{cases}
\end{equation*}
Let $w$ be such that:
\begin{equation*}
w =
\begin{cases}
0 & x \geq y\\
y - x & otherwise
\end{cases}
\end{equation*}
We can express $w$ by using \eqref{eq:noncontinu1}
and~\eqref{eq:noncontinu2}.

\item Minimums: Expressing minimums is based on the same idea than
  expressing maximums:
\begin{equation*}
z  =  min(x, y)
   = x - (x - y)
\begin{cases}
0 & x \leq y\\
x - y & otherwise
\end{cases}
\end{equation*}
We do not detail how minimums are expressed, since
it is done the same way as maximums.



\item Expressing several conditions: In definitions
  like~\eqref{eq:dij}, several conditions can force the value of a variable.
\begin{equation*}
w = 
\begin{cases}
0 & x \leq y  \hspace{0.1cm}or \hspace{0.1cm} z \geq u\\
0 & otherwise
\end{cases}
\end{equation*}
Translating such definitions into inequalities requires the use of one
binary variable for each condition and one binary variable to express
the ``or''.

Let $bin1$, $bin2$ be such that:
$\displaystyle bin1 =\left\{%
\begin{array}{ccl }
1 & if & z - u \geq 0 \\ 0 && otherwise
\end{array}\right.$
and
$\displaystyle bin2 =\left\{%
\begin{array}{ccl }
1 & if &  x - y \leq 0 \\ 0 && otherwise
\end{array}\right.$

These two definitions can be expressed using
\eqref{eq:bool1} and~\eqref{eq:bool2}.

Finally $bin3$ is a binary variable which is equal to $1$ if $bin1$ or
$bin2$ are equal to $1$ and $0$ otherwise:
\begin{equation}
bin3 =
\begin{cases}
1 & bin1 + bin2 \geq 1\\
0 & otherwise
\end{cases}
\label{eq:rho}
\end{equation}
Since $bin1$, $bin2$ and $bin3$ are binary variables,
\eqref{eq:rho} can be easily expressed as:
\begin{eqnarray}
bin1 & \leq & bin3 \label{eq:bool6}\\
bin2 & \leq & bin3\label{eq:bool7}\\
bin3 & \leq & bin1 + bin2\label{eq:bool8}
\end{eqnarray}
Thus, when $bin1$ and $bin2$ are $0$, \eqref{eq:bool8} forces
$bin3$ to be $0$ whereas when $bin1$ or $bin2$ are equal to $1$,
\eqref{eq:bool6} and~\ref{eq:bool7} forces $bin3$ to be
equal to $1$.

\end{enumerate}

\subsection{Objective function}
Minimizing the energy consumption of a program described as a set of
tasks is the objective function of the linear programming formulations
described above. For a task $T_i$ with a power consumption at a
frequency $f$, $P_i^f$ and executed at frequency $f$ during $tT_i^f$,
the energy consumption of the whole program for its whole execution
time is:

\begin{equation*}
min (\sum\limits_{T_i} (\sum\limits_f (tT_i^f \times P_i^f)))
\end{equation*}

\subsection{Task constraints}
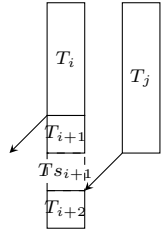
\begin{figure}[h]
\centering
\begin{tikzpicture}
\draw (0,3) rectangle (0.5,1.5) node [midway,font=\scriptsize]{$T_i$};
\draw (0,1.5) rectangle (0.5,1) node [midway,font=\scriptsize]{$T_{i+1}$};
\draw (0,1) [dashed] rectangle (0.5,0.5) node [midway,font=\scriptsize]{$Ts_{i+1}$};
\draw (0,0.5) rectangle (0.5,0) node [midway,font=\scriptsize]{$T_{i+2}$};

\draw (1,3) rectangle (1.5,1) node [midway,font=\scriptsize]{$T_j$};
\draw [->,>=stealth] (1,1) -- (0.5,0.5);
\draw [->,>=stealth] (0,1.5) -- (-0.5,1);

\end{tikzpicture}
\caption{Task configuration}\label{fig:appendix_example}
\end{figure}

Let $T_i, T_{i+1}, T_{i+2}, T_j$ be four tasks such that: $T_i,
T_{i+1}, T_{i+2}$ are consecutive and on the same processor. $T_i$
ends with a message sending creating $T_{i+1}$ which ends with a
reception from $T_j$ which generates $T_{i+2}$ as shown in
Figure~\ref{fig:appendix_example}.

\begin{eqnarray*}
eT_i & = & bT_i + \sum\limits_{f} tT_i^f \\
\sum\limits_{f} \delta_i^f & = & 1 \\
bT_{i+1} & = & eT_{i}\\
bTs_{i+1} & = & eT_{i+1}\\
eTs_{i+1} & \geq & eT_{j} + M_j^{i+1}\\
eTs_{i+1} & \geq & bTs_{i+1}\\
bT_{i+2} & = & eTs_{i+1}\\
tT_i^f & = &\delta_i^f \times execT_i^f
\end{eqnarray*}

\subsection{Workload approach}
\subsubsection{Additional variable}
\begin{tabular}{lcl}
$\gamma_i$ &:& A binary variable used to say if a workload duration is $0$ or not\\
$M$ &:& A large constant
\end{tabular}

\begin{center}
$\begin{array}{lcccl}
bW_i & \geq & bT_j &&\\
eW_i & \leq & eT_j&&\\
tT_i^f & = & \sum\limits_{\substack{W_j \\ T_i\in W_j}} tW_j^f&&\\ 
dW_i & = & \sum\limits_{f} tW_i^f&&\\ 
\end{array}$
\end{center}
Using \eqref{eq:bool1}, \eqref{eq:bool2}
and~\eqref{eq:tWifTh}, we express definition~\eqref{eq:twifB} as:
\begin{center}
$\begin{array}{lcccl} 
    tW_i^f & \geq & Th \times \overline{tW_i^f}&&\\ 
    tW_i^f & \leq & M \times \overline{tW_i^f}&&\\ 
\end{array}$
\end{center}
Using \eqref{eq:bool1}, \eqref{eq:bool2}, \eqref{eq:noncontinu1}
and~\eqref{eq:noncontinu2} and $\gamma_i$ as the binary variable, we
express definition~\eqref{eq:dWi} as:
\begin{center}
$\begin{array}{lcccl} 
    eW_i - bW_i & \leq & M\times\gamma_i &&\\ 
    bW_i - eW_i & \leq & M\times(1-\gamma_i) & , \gamma_i \in \{0,1\}&\\ 
    eW_i - bW_i & \leq & dW_i & \leq & M\times\gamma_i \\
    eW_i - bW_i + dW_i &\leq & 2\times(eW_i - bW_i) + M\times(1-\gamma_i)&&\\
\end{array}$
\end{center}

\subsubsection{Proof of workload duration}\label{ssec:workload_proof}
We want to proove that if two workloads $W$ and $W'$ are possible, but
they violate the precedence constraint between the tasks, then the
duration of at least one of them is zero. We provide the proof for
workloads with a cardinality equals to $2$ since the proof remains the
same for larger workloads.

Let $W=(T_i, T_j)$ and $W'=(T_i', T_j')$ such that $T_i$ preceeds
$T_i'$ and $T_j'$ preceeds $T_j$. We want to prove that $dW=0$ or
$dW'=0$.

\begin{lemma}
Let $W=(T_i, T_j)$ and $W'=(T_i', T_j')$. If $bT_i' \geq eT_i$ and
$bT_j \geq eT_i'$, then $dW = 0$ or $dW'=0$.
\end{lemma}\label{lemma:workload}

\begin{proof}
Let us proove lemma~\ref{lemma:workload} by contradiction. Let us assume
that $dW \neq 0$ and $dW' \neq 0$.

From definition~\eqref{eq:twifB}:
$\begin{array}{lcl}
dW \neq 0 &\Leftrightarrow& eW \geq bW\\
dW' \neq 0 &\Leftrightarrow& eW' \geq bW'
\end{array}$

From constraints~\eqref{eq:workload_b}
and~\eqref{eq:workload_e}:

\begin{minipage}{0.5\linewidth}
\begin{align}
bW &\geq bT_i \nonumber\\
bW &\geq bT_j \nonumber\\
eW &\leq eT_i\label{eq:workend}\\
eW &\leq eT_j\nonumber
\end{align}
\end{minipage}
and
\begin{minipage}{0.5\linewidth}
\begin{align*}
bW' &\geq bT_i'\\
bW' &\geq bT_j'\\
eW' &\leq eT_i'\\
eW' &\leq eT_j'
\end{align*}
\end{minipage}



But $bT_i' \geq eT_i$ and $bT_j \geq eT_i'$, thus:
\begin{eqnarray}
bW \geq bT_j \geq eT_j' \geq eW'\label{eq:workload1}\\
bW' \geq bT_i' \geq eT_i \geq eW\label{eq:workload2}
\end{eqnarray}
If we consider \eqref{eq:workload1},~\eqref{eq:workload2}
and~\eqref{eq:workend}:
\begin{equation*}
bW' \geq bT_i' \geq eT_i \geq bW \geq eW'
\end{equation*}
Thus $bW' \geq eW'$ which by definition~\eqref{eq:twifB} implies that
$dW' = 0$ which leads to a contradiction.
\end{proof}

\subsection{Frequency switch approach}
Note that we do not detail how the threshold condition is handled
since it is done the same as for the workloads.

\subsubsection{Additional variables}
\begin{tabular}{lcl}
$\zeta_{ip}^f$ &:& A binary variable used to say if a workload is executed at a frequency $f$ or not\\
$y_{ij}^f$ &:& The maximum between $bT_i$ and $c_{jp}^f$\\
$w_{ij}^f$ &:& A variable used to express $y_{ij}^f$. It is equal to $0$ if $bT_i$ is the maximum, and $c_{jp}^f - bT_i$ otherwise\\
$\alpha_{ij}^f$ &:& A binary variable used to verify whether $bT_i \geq c_{jp}^f$\\
$z_{ij}^f$ &:& The minimum between $eT_i$ and $c_{\{j+1\}p}^f$\\
$g_{ij}^f$ &:& A variable used to express $z_{ij}^f$. It is equal to $0$ if $eT_i$ is the minimum, and $eT_i - c_{\{j+1\}p}^f$ otherwise\\
$\beta_{ij}^f$ &:& A binary variable used to verify whether $eT_i \leq c_{\{j+1\}p}^f$\\
$\psi_{ij}^f$ &:& A binary variable used to check if $bT_i - c_{\{i+1\}p}^f \geq 0$\\
$\phi_{ij}^f$ &:& A binary variable used to check if $eT_i - c_{ip}^f \leq 0$\\
$\rho_{ij}^f$ &:& A binary variable used to check if $\psi_{ij}^f$ or $\phi_{ij}^f$ are true\\
$M$ &:& A large constant
\end{tabular}

\subsubsection{Constraints}
\begin{center}
$\begin{array}{lcccl}
c_{\{i+1\}p}^{f'} & \geq & c_{ip}^f &&\\
c_{\{i+1\}p}^{f'} - c_{ip}^f & \geq & Th \times \zeta_{ip}^f&&\\
c_{\{i+1\}p}^{f'} - c_{ip}^f & \leq & M \times \zeta_{ip}^f &&\\
tT_i^f & = & \sum\limits_{j} d_{ij}^f&&\\
\end{array}$
\end{center}
Expressing definition~\eqref{eq:dij} 
as inequalities requires the use of \eqref{eq:noncontinu1}
and~\eqref{eq:noncontinu2} for the maximum and the minimum such that:
\begin{center}
$\begin{array}{lcl}
y_{ij}^f & = & max (bT_i, c_{jp}^f) \\
        & = & bT_i + w_{ij}^f
\end{array}$
such that: $\displaystyle w_{ij}^f =\left\{%
\begin{array}{ccl }
0 & if & bT_i \text{ is the maximum} \\c_{jp}^f - bT_i && otherwise
\end{array}\right.$
\end{center}
\begin{center}
$\begin{array}{lcl}
z_{ij}^f & = & min (eT_i, c_{\{j+1\}p}^{f'}) \\
        & = & eT_i - g_{ij}^f
\end{array}$
such that: $\displaystyle g_{ij}^f =\left\{%
\begin{array}{ccl }
0 & if & eT_i \text{ is the minimum} \\c_{\{j+1\}p}^f - eT_i && otherwise
\end{array}\right.$
\end{center}
Let $\alpha_{ij}^f$ be the binary variable used for the maximum and
$\beta_{ij}^f$ the one used for the minimum. By replacing the
corresponding variables in \eqref{eq:noncontinu1}
and~\eqref{eq:noncontinu2}, we obtain the following inequalities for
the maximum:
\begin{center}
$\begin{array}{lcccl}
c_{jp}^f - bT_i & \leq & M \times\alpha_{ij}^f&&\\
bT_i - c_{jp}^f & \leq & M\times(1-\alpha_{ij}^f) &,\alpha_{ij}^f \in \{0,1\} &\\
c_{jp}^f - bT_i &\leq & w_{ij}^f & \leq & M\times\alpha_{ij}^f \\
c_{jp}^f - bT_i + w_{ij}^f &\leq & 2\times(c_{jp}^f - bT_i) + M\times(1-\alpha_{ij}^f)&&\\
\end{array}$
\end{center}
and the following for the minimum:
\begin{center}
$\begin{array}{lcccl}
eT_i - c_{\{j+1\}p}^{f'} & \leq & M \times\beta_{ij}^f&&\\
c_{\{j+1\}p}^{f'} - eT_i& \leq & M\times(1-\beta_{ij}^f)&,\beta_{ij}^f \in \{0,1\} &\\
eT_i - c_{\{j+1\}p}^{f'} &\leq & g_{ij}^f & \leq & M\times\beta_{ij}^f \\
eT_i - c_{\{j+1\}p}^{f'} + g_{ij}^f &\leq & 2\times(eT_i - c_{\{j+1\}p}^{f'}) + M\times(1-\beta_{ij}^f)&&\\
\end{array}$
\end{center}
Finally, using \eqref{eq:bool6}, \eqref{eq:bool7}
and~\eqref{eq:bool8} and the binary variables $\psi_{ij}^f$,
$\phi_{ij}^f$ and $\rho_{ij}^f$ as $bin1$, $bin2$ and $bin3$
respectively and using \eqref{eq:noncontinu1}
and~\eqref{eq:noncontinu2}, $d_{ij}$ can be expressed as:
\begin{center}
$\begin{array}{lcccl}
\phi_{ij}^f & \leq & \rho_{ij}^f &&\\
\psi_{ij}^f & \leq & \rho_{ij}^f &&\\
\rho_{ij}^f & \leq & \phi_{ij}^f + \psi_{ij}^f&&\\
z_{ij}^f - y_{ij}^f  &\leq & d_{ij}^f & \leq & M\times(1-\rho_{ij}^f) \\
z_{ij}^f - y_{ij}^f + d_{ij}^f &\leq & 2\times(z_{ij}^f - y_{ij}^f) + M\times\rho_{ij}^f&&\\
\end{array}$
\end{center}

\bibliographystyle{abbrv}
\bibliography{sc2014}

\end{document}